\documentclass[10pt,sigconf,authorversion,letterpaper]{acmart}

\setcopyright{acmlicensed}

 \usepackage{todonotes}
\AtBeginDocument{%
  \providecommand\BibTeX{{%
    \normalfont B\kern-0.5em{\scshape i\kern-0.25em b}\kern-0.8em\TeX}}}

\copyrightyear{2021} 
\acmYear{2021} 
\setcopyright{acmcopyright}\acmConference[ANRW '21]{Applied Networking Research Workshop}{July 24--30, 2021}{Virtual Event, USA}
\acmBooktitle{Applied Networking Research Workshop (ANRW '21), July 24--30, 2021, Virtual Event, USA}
\acmPrice{15.00}
\acmDOI{10.1145/3472305.3472322}
\acmISBN{978-1-4503-8618-0/21/07}

\begin{document}

\title{CCID5: An implementation of the BBR Congestion Control algorithm for DCCP and its impact over multi-path scenarios}


\author{Nathalie Romo Moreno}
\author{Markus Amend}
\affiliation{%
  \institution{Deutsche Telekom}
  \streetaddress{Deutsche-Telekom-Allee 9}
  \city{Darmstadt}
  \country{Germany}}
\email{nathalie.romo-moreno@telekom.de}
\email{markus.amend@telekom.de}

\author{Anna Brunstrom}
\author{Andreas Kassler}
\affiliation{%
  \institution{Karlstads Universitet}
  \streetaddress{P.O. Box 1212}
  \city{Karlstad}
  \country{Sweden}
  \postcode{43017-6221}
}
\email{anna.brunstrom@kau.se}
\email{andreas.kassler@kau.se}

\author{Veselin~Rakocevic}
\affiliation{%
  \institution{City, University of London}
  \streetaddress{Northampton Square}
  \city{London}
  \country{United Kingdom}
  \postcode{EC1V 0HB}
}
\email{Veselin.Rakocevic.1@city.ac.uk}

\renewcommand{\shortauthors}{Moreno and Amend, et al.}

\begin{abstract}
Providing multi-connectivity services is an important goal for next generation wireless networks, where multiple access networks are available and need to be integrated into a coherent solution that efficiently supports both reliable and non reliable traffic. Based on virtual network interfaces and per path congestion controlled tunnels, the MP-DCCP based  multiaccess aggregation framework presents a novel solution that flexibly supports different path schedulers and congestion control algorithms as well as reordering modules. The framework has been implemented within the Linux kernel space and has been tested over different prototypes. Experimental results have shown that the overall performance strongly depends upon  the congestion control algorithm used on the individual DCCP tunnels, denoted as CCID. In this paper, we present an implementation of the BBR (Bottleneck Bandwidth Round Trip propagation time) congestion control algorithm for DCCP in the Linux kernel. We show how BBR  is integrated into the MP-DCCP multi-access framework and evaluate its performance over both single and multi-path environments. Our evaluation results show that BBR improves the performance compared to CCID2 (TCP-like Congestion Control) for multi-path scenarios due to the faster response to changes in the available bandwidth, which reduces latency and increases performance, especially for unreliable traffic. The MP-DCCP framework code, including the new CCID5 is available as OpenSource\footnote{\url{https://github.com/telekom/mp-dccp/tree/d9b4b7471a69184105885abbc4d45c011b82543d/net/dccp/ccids}}.
\end{abstract}




\maketitle

\section{Introduction}
Multi-connectivity and Hybrid Access technologies have become an active research topic during the last years, as the possibility of simultaneously using heterogeneous networks brings along a series of advantages that are appealing for service and network operators. Capacity aggregation, reliability and flexible resource management are key aspects for multi-access solutions aimed to flexibly improve the performance of applications having diverse demands in terms of end-to-end latency, throughput and reliability. Driven by that interest, several solutions have been developed and are under standardization, including the Broadband Forum standard for Hybrid Access~\cite{bb_forum}\cite{bbf-tr470} and the Access Traffic Steering Switching Splitting (ATSSS) specified first in 3GPP Rel. 16~\cite{3gpp.23.501}. Within the ATSSS architecture,  MPTCP~\cite{RFC8684} has already been defined as a solution for traffic splitting, with the constraint of being limited to the transport of reliable and strict in-order traffic.

However, several applications such as voice and video streaming have relaxed constraints on end-to-end reliability and using MPTCP within such multi-access frameworks may lead to head-of-line blocking when operated over multiple paths with different latency and capacity characteristics. Consequently, MP-DCCP~\cite{mpdccp_proto} has emerged, a protocol that extends DCCP's~\cite{rfc4340} connection-oriented transport layer for congestion-controlled but unreliable data transmission to support multi-path sessions. In addition, a multi-path access traffic aggregation framework based on MP-DCCP has been designed and implemented. The framework combines the multi-path protocol with other elements such as virtual network interfaces, flexible packet scheduling and reordering functions to enable the transparent transport of any IP traffic across multiple access networks~\cite{mpdccp_frame}. 

The MP-DCCP based framework has been evaluated with different physical prototypes and simulation environments~\cite{frame_eval}. Experimental results show that a small latency difference among the different paths improves the end-to-end performance, and that this improvement is strongly dependent on the congestion control (CC) algorithm used. DCCP provides a framework for adding new congestion control algorithms. The currently standardized CC algorithms for DCCP, which are denoted by a CCID (Congestion Control Identifier), correspond to CCID2~\cite{rfc4341}, CCID3~\cite{rfc4342} and CCID4~\cite{rfc5622}. All of them are based on existing TCP loss-based CC algorithms. 

In this paper we aim to improve the MP-DCCP framework performance by extending  DCCP with a new congestion control mechanism that aims to provide more efficient control  and maintain low latency while still being able to support a high sending rate. We implement a new CCID for DCCP, identified as CCID5, that leverages  BBR~\cite{bbr1} to estimate the Bottleneck Bandwidth (BtlBW) and the Round Trip propagation time(RTProp). We describe the details of the BBR implementation for DCCP. We evaluate our implementation in a multipath testbed using the MP-DCCP framework and measure end-to-end latency as well as aggregated throughput over multiple paths. Our results show that using CCID5 it is possible to maintain a lower latency due to better control of the bottleneck buffer, significantly reducing bufferbloat on the individual paths. In addition, in scenarios where path characteristics change, CCID5 provides a faster response, which enables the MP-DCCP framework to better aggregate capacity.

The rest of the paper is structured as follows. In section 2, we provide an overview of our CCID5 implementation of BBR and show how it is integrated into the MP-DCCP access aggregation framework. Section 3 introduces our evaluation setup and metrics. In Section 4, we bring experimental results and their analysis. The paper concludes with ideas for future work in section 5.

\section{Design and implementation}\label{impl}
\begin{figure}[h]
	\centering
	\def\svgwidth{\columnwidth}
\begingroup%
  \makeatletter%
  \providecommand\color[2][]{%
    \errmessage{(Inkscape) Color is used for the text in Inkscape, but the package 'color.sty' is not loaded}%
    \renewcommand\color[2][]{}%
  }%
  \providecommand\transparent[1]{%
    \errmessage{(Inkscape) Transparency is used (non-zero) for the text in Inkscape, but the package 'transparent.sty' is not loaded}%
    \renewcommand\transparent[1]{}%
  }%
  \providecommand\rotatebox[2]{#2}%
  \newcommand*\fsize{\dimexpr\f@size pt\relax}%
  \newcommand*\lineheight[1]{\fontsize{7}{7}\selectfont}%
  \ifx\svgwidth\undefined%
    \setlength{\unitlength}{127.5000036bp}%
    \ifx\svgscale\undefined%
      \relax%
    \else%
      \setlength{\unitlength}{\unitlength * \real{\svgscale}}%
    \fi%
  \else%
    \setlength{\unitlength}{\svgwidth}%
  \fi%
  \global\let\svgwidth\undefined%
  \global\let\svgscale\undefined%
  \makeatother%
  \begin{picture}(1,0.33970192)%
    \lineheight{1}%
    \setlength\tabcolsep{0pt}%
    \put(0,0){\includegraphics[width=\unitlength,page=1]{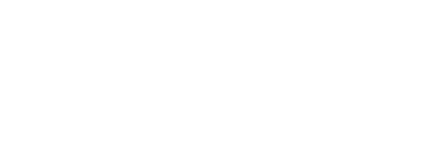}}%
    \put(0.50588232,0.31792937){\color[rgb]{0,0,0}\makebox(0,0)[t]{\lineheight{1.25}\smash{\begin{tabular}[t]{c}Startup\end{tabular}}}}%
    \put(-0.4307548,0.54972889){\color[rgb]{0,0,0}\makebox(0,0)[lt]{\begin{minipage}{1.11162572\unitlength}\raggedright \end{minipage}}}%
    \put(0,0){\includegraphics[width=\unitlength,page=2]{bbr_state_machine.pdf}}%
    \put(0.45294134,0.12969415){\color[rgb]{0,0,0}\makebox(0,0)[lt]{\lineheight{1.25}\smash{\begin{tabular}[t]{l}ProbeBW\end{tabular}}}}%
    \put(0,0){\includegraphics[width=\unitlength,page=3]{bbr_state_machine.pdf}}%
    \put(0.45882362,0.21792953){\color[rgb]{0,0,0}\makebox(0,0)[lt]{\lineheight{1.25}\smash{\begin{tabular}[t]{l}Drain\end{tabular}}}}%
    \put(0.44117663,0.00028235){\color[rgb]{0,0,0}\makebox(0,0)[lt]{\lineheight{1.25}\smash{\begin{tabular}[t]{l}ProbeRTT\end{tabular}}}}%
  \end{picture}%
\endgroup%

	\caption{BBR state machine}
	\label{fig:mp-bbr-state}
	\Description{BBR state machine}
\end{figure}

The choice of BBR as the new CCID for (MP-)DCCP was made based on the fact that its implementation in TCP  has proven to overcome problems like low throughput, long delays, and buffer-bloating typically present in loss-based CC algorithms~\cite{bbr1}. In more detail, BBR aims to balance low latency while maintaining a high sending rate to reach an optimal point of operation by fulfilling two conditions:
\begin{itemize}
	\item The amount of data in flight must be equal to the Bandwidth Delay Product (BDP), where
\end{itemize}
\begin{equation}
	BDP = BtlBw*RTprop
\end{equation}
\begin{itemize}
	\item The bottleneck packet arrival must match the $BtlBw$ to ensure its full utilization.
\end{itemize}
BBR estimates the values of $BtlBw$ and $RTprop$ to later use them as input in the calculation of three control variables: $cwnd$ (congestion window), $pacing\_rate$ and $send\_quantum$.  The way the control parameters are updated is governed by the BBR state machine illustrated in Figure~\ref{fig:mp-bbr-state}. In the initial $Startup$ state, the sending rate will increase rapidly until the pipe is detected to be full. Afterwards, the data rate will be reduced so any possible queue can be drained, to finally enter into the $ProbeBW$ state, where the amount of data in flight is slightly increased to probe for more possible bandwidth available. From any of these states, the algorithm can jump into the $ProbeRTT$ phase. Here the data inflight is reduced to probe for lower RTTs. Each state defines specific values for two dynamic gains: \textit{cwnd\_gain} and \textit{pacing\_gain}, which will finally be used in the calculation of the aforementioned control variables. The dynamic gain values for each stage and the calculation process of the control variables can be found in ~\cite{bbr2}.

The existing implementation of the MP-DCCP multi-path framework, including the new CCID5, is deployed within the Linux kernel under its 4.14.111 version. We implemented the core functionalities of the BBR algorithm  on its version 1 (BBRv1), such as the state machine and the path capacity estimation following the guidelines and pseudo code depicted in~\cite{bbr2} and~\cite{bbr3} respectively, as well as the code of the Linux kernel implementation of BBR for TCP. Although BBRv2 is already available and promises to overcome the fairness problems existing on BBRv1, we decided to start with the implementation of version 1 with the aim of confirming the applicability of the BBR conceptual basis to DCCP. Furthermore, under the Hybrid Access scenario, where the algorithm is intended to be tested, fairness doesn't represent a major concern, as dedicated physical resources are available (e.g. LTE and DSL links).
\subsection{TCP vs DCCP implementation}
\begin{figure}[h]
	\centering
	\def\svgwidth{\columnwidth}
	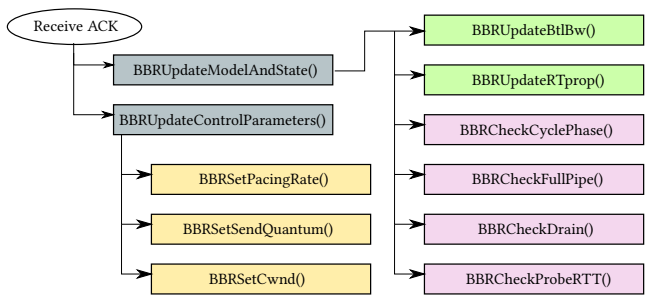
	\caption{BBR functions executed upon ACK received}
	\label{fig:mp-bbr}
	\Description{BBR functions executed upon ACK received}
\end{figure}

\begin{figure}[h]
	\centering
	\includegraphics[width=\linewidth]{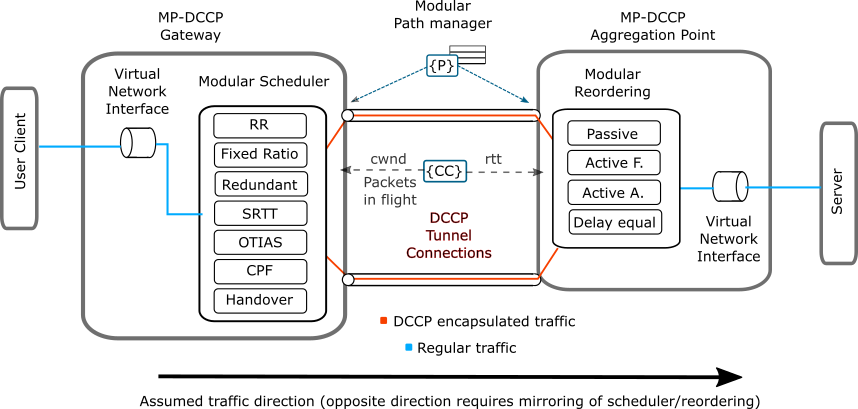}
	\caption{MP-DCCP Multi-path framework architecture}
	\label{fig:mp-frame}
	\Description{MP-DCCP Multi-path framework architecture}
\end{figure}

The CCID5 implementation reuses the BBR-TCP code functions which are in charge of the path estimation, the state machine transitions and the update of the control parameters, as illustrated in the function scheme of Figure~\ref{fig:mp-bbr}, but it also implements additional functionalities to fulfill the DCCP specification requirements.
The DCCP standard~\cite{rfc4340} defines the implementation of Congestion Control in a modular manner, thus, in addition to the description of how data packets are congestion controlled, each CCID must specify the format of the ACK packets, the timing of their generation, and how they are congestion controlled. In the case of TCP, the definition of the ACK format and its generation are part of the protocol itself and not of the CC algorithm, and the functionality of Congestion Control for ACKs is nonexistent. For the CCID5 implementation, those missing definitions were taken from the CCID2 profile, hence, CCID5 uses the same ACK format as CCID2, including ACK vectors containing the same information that can be found in SACK options, and implements the ACK ratio as ACK Congestion Control mechanism. In addition, the different variables and functions used to track packets in flight, packets acknowledged, and their corresponding sending and arrival times as well as the function to detect application-limited periods were replicated from the CCID2 implementation, thus, the values of the $cwnd$ and the amount of data delivered used in the $BtlBw$ estimation are maintained in packet units.  
Finally, the scheduling and reordering engines of the multi-path framework require information about the congestion status and the RTT estimation of the available paths (see Figure ~\ref{fig:mp-frame}). The CCID5 module exposes this information towards the MP-DCCP framework using a suitable interface. 
\section{Evaluation Scenarios and Metrics}
We use the following two scenarios to evaluate the performance of the new CCID5. To validate whether the BBR approach and algorithm (initially designed for TCP) are applicable for DCCP as well, we use a single path topology and compare the response in both protocols. The second scenario aims to evaluate the performance of CCID5 in a multi-access aggregation scenario comparing CCID5 with CCID2. We measure the received throughput and the end-to-end latency. We use \texttt{Iperf3}~\cite{iperf3} to generate UDP and TCP traffic.  DCCP traffic is generated using a patched version of \texttt{Iperf}~\cite{iperf_patch}.
\subsection{Evaluation Metrics}
We used the following methodology to collect evaluation metrics:
\begin{itemize}
	\item End to end latency: The incoming traffic on server side is mirrored and forwarded back to the client. On the client side, the traffic transmitted and received is captured separately by using Wireshark. We postprocess the capture files to extract sequence numbers and timestamp information to calculate per-packet latency. By mirroring the received traffic back to the sender, we avoid the problem of clock synchronization. The additional path used to mirror the traffic from server to client might add a latency offset lower than 1ms, however, that offset is neglectable as the measurements are taken for comparison purposes and all of them would be affected equally.
\end{itemize}
\begin{itemize}
	\item Received throughput: Incoming traffic at server side is captured. We use Wireshark's IP graph tool to plot throughput over time.
\end{itemize}
\subsection{Single Path scenario}
\begin{figure}[h]
	\centering
	\includegraphics[width=\linewidth]{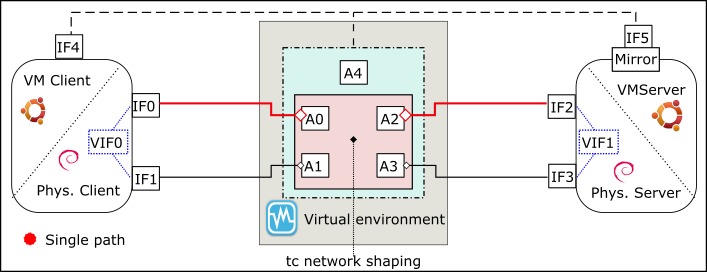}
	\caption{Test topology}
	\label{fig:topo_all}
	\Description{Two Virtual Machines connected via host adapters using a single path}
\end{figure}

We set up the topology from Figure ~\ref{fig:topo_all} using a virtual environment deployed over a host machine running Ubuntu 16.04.6 and Oracle VM VirtualBox version 5.1.38. We interconnected  two virtual machines (VMs) trough host only adapters (denoted as Ax). The VMs run the same Ubuntu version as the host but with the Linux Kernel version 4.14.114 enhanced with the CCID5 module. In this scenario only the path depicted in red is used to transmit the traffic.
\subsubsection{Experiment 1: Bandwidth change response}\label{sp-exp1}To evaluate the responsiveness of the CCs to changes in available capacity, we send TCP and DCCP flows from VM client to VM server. The available bandwidth is reduced to 10Mbps during the test by applying \texttt{tc} rules on the host adapters A0 and A1. Test parameters are listed in Table ~\ref{tab:single_bwch}
\begin{table}[h]
	\caption{Single-path BW change test}
	\label{tab:single_bwch}
	\begin{tabular}{ccccc}
		\toprule
		\begin{tabular}{c}Traffic\\type\end{tabular}&\begin{tabular}{c}CC\\algorithm\end{tabular} & \begin{tabular}{c}Tx\\rate\end{tabular}&\begin{tabular}{c}Path\\BW\end{tabular}&Duration\\
		\midrule
		\begin{tabular}{c}TCP\\\\DCCP\end{tabular}&\begin{tabular}{c}CUBIC\\BBR\\CCID2\\CCID5\end{tabular}&15Mbps&\begin{tabular}{c}1G\\for $t < 5s$\\10Mbps\\for $t \ge 5s$\end{tabular}&10s\\
		\bottomrule
	\end{tabular}
\end{table}
\subsection{Multi-Path scenario}
To evaluate the impact of CCID5 on the performance in a multi-access traffic aggregation scenario, we compare CCID5 with CCID2. To validate  the performance in a realistic access bundling environment capable to transport non-reliable traffic, we use the MP-DCCP framework with the different CCIDs to tunnel application traffic using both UDP and TCP flows.
 As mentioned in section~\ref{impl} the multi-path framework provides modular scheduling and reordering functions with different algorithms available. During these experiments we use Cheapest Pipe First (CPF) and Round Robin (RR) schedulers in combination with Passive and Active-Fixed reordering.
\begin{itemize}
	\item CPF scheduler: Allocates packets based on a predefined path priority. Paths fully congested are skipped from the selection.
	\item RR scheduler: Alternates packet-sending through all the  available paths. Paths fully congested are skipped from the selection.
	\item Passive reordering: Arriving packets are directly forwarded to the virtual interface without further processing.
	\item Active-Fixed: Reads packet sequence numbers to verify in order arrival. When a gap is detected, a buffer is used to store received packets until the missing one(s) arrive, or a fixed timer expires.
\end{itemize}
\subsubsection{Experiment 1: UDP over Multi Path DCCP}\label{mp-exp1}
We use the same virtual topology depicted in Figure ~\ref{fig:topo_all} with the two paths available and the virtual interfaces. In this case we follow the same procedure as in the single-path scenario, we start a test under stable conditions, and after 5 seconds we limit the bandwidth available in both paths using \texttt{tc}. Table ~\ref{tab:multi_udp_test2} details the experimental settings.
\begin{table}[h]
	\caption{UDP over MP-DCCP test cases}
	\label{tab:multi_udp_test2}
	\begin{tabular}{ccccc}
		\toprule
        Sched. & Reordering & Tx rate & Per-path BW & Duration\\
		\midrule
		CPF & Active Fixed & 15Mbps & \begin{tabular}{c}1G \\for $t < 5s$\\10Mbps\\ for $t \ge 5s$\end{tabular} & 20s\\
		\bottomrule
	\end{tabular}
\end{table}
\subsubsection{Experiment 2: TCP over Multi-Path DCCP} 
In this experiment, we evaluate the performance when transporting TCP traffic. Although the MP-DCCP framework was initially developed to fill the need of a multi-connectivity solution with support for non reliable traffic, it is also capable of aggregating any traffic above layer 3. Furthermore, testing CCID5 in a scenario with nested congestion control can give insights on how the algorithm would behave in the case of transporting QUIC over the MP-DCCP solution.

In contrast to the previous tests, we do not manually set the sending rate of the application but rather let TCP's congestion control decide it. The test topology corresponds to Figure~\ref{fig:topo_all}, but this time deployed over  two PC engines APU boards running Debian 9.3 directly connected (host adapters are not part of this scenario). We limit the interface speed for data traffic using ethtool to avoid CPU bottlenecks impacting our results. We detail the scenario setup in Table ~\ref{tab:multi_tcp_test1}.
\begin{table}[h]
	\caption{Multi-path TCP test case}
	\label{tab:multi_tcp_test1}
	\begin{tabular}{ccccc}
		\toprule
        Sched. & Reordering & Tx Rate &  Per-path BW & Duration\\
		\midrule
        RR & Active-Fixed & NA &  10Mbps & 20s\\
		\bottomrule
	\end{tabular}
\end{table}
\section{Experimental Results}
\subsection{Single path scenario}
\begin{figure}[h]
	\centering
	\includegraphics[width=\linewidth]{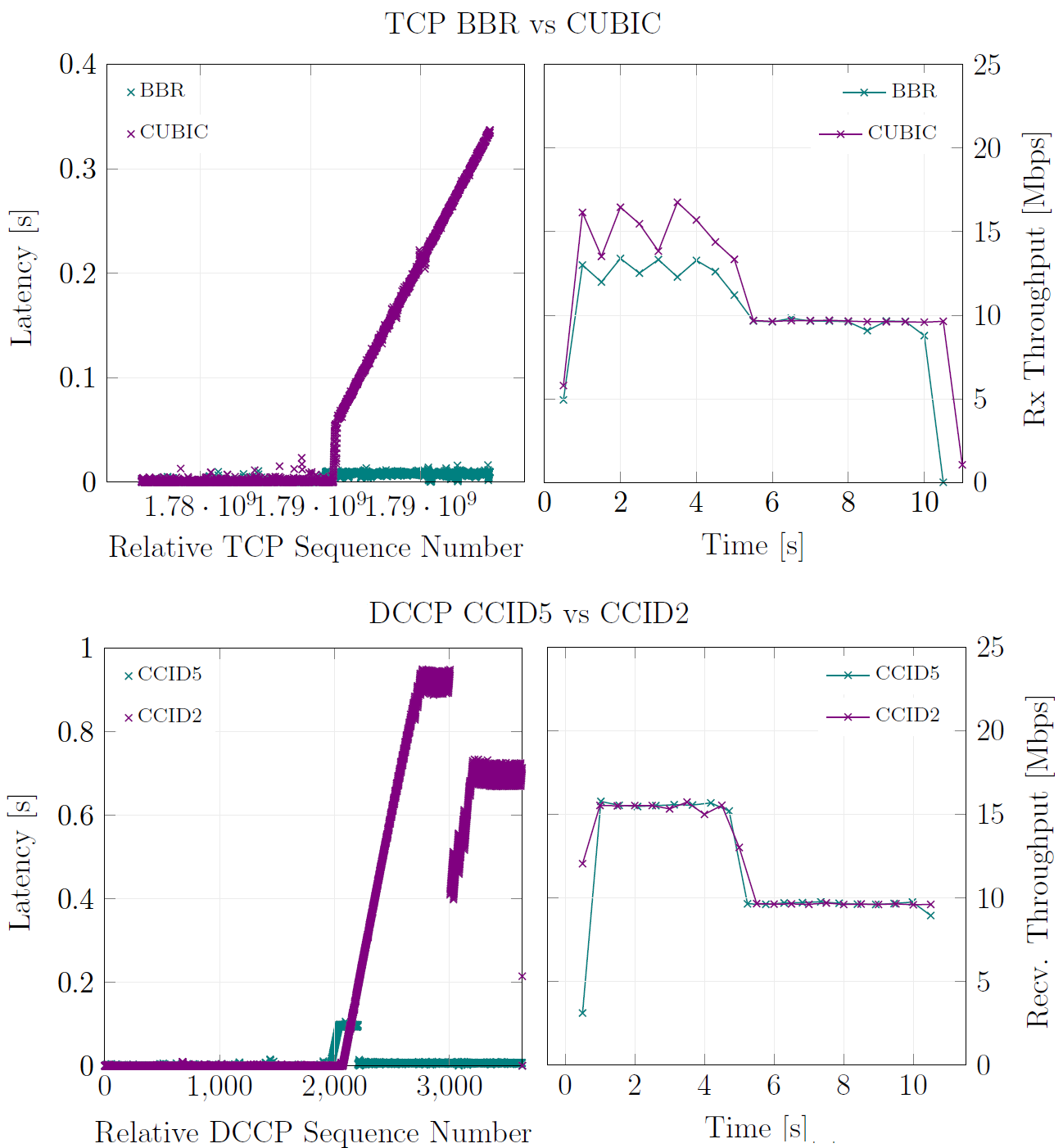}
	\caption{Latency and Rx throughput comparison - BBR vs CUBIC - CCID vs CCID5.}
	\label{fig:sp-TCP-bwch}
	\Description{Latency and Rx throughput comparison for BBR and CUBIC.}
\end{figure}

Figure ~\ref{fig:sp-TCP-bwch} shows the results for the "bandwidth-change response" experiment described in ~\ref{sp-exp1} for TCP and DCCP. The reaction to the Bandwidth limitation imposed in the path is clear, the received throughput is immediately limited to the value set by the \texttt{tc} configuration in the host machine. In addition, as \texttt{tc} uses queuing disciplines to limit the required bandwidth, a new buffer is created in the path, the senders initially try to keep the sending rate imposed by the application, which causes the buffer to fill up and the packets to be delayed. Both BBR and CCID5 manage to fully utilize the bandwidth available while avoiding buffer bloat and therefore,  keeping the values of the latencies low.  CUBIC and CCID2 fill up the buffer causing bigger latencies and even packet loss in the case of CCID2, which can be observed in the drop of the latency values almost at the end of the test, given that the amount of data inflight gets reduced by the cut down of the congestion window after the loss detection. 
\subsection{Multi path scenario - UDP traffic}
\begin{figure}[h]
	\centering
	\includegraphics[width=\linewidth]{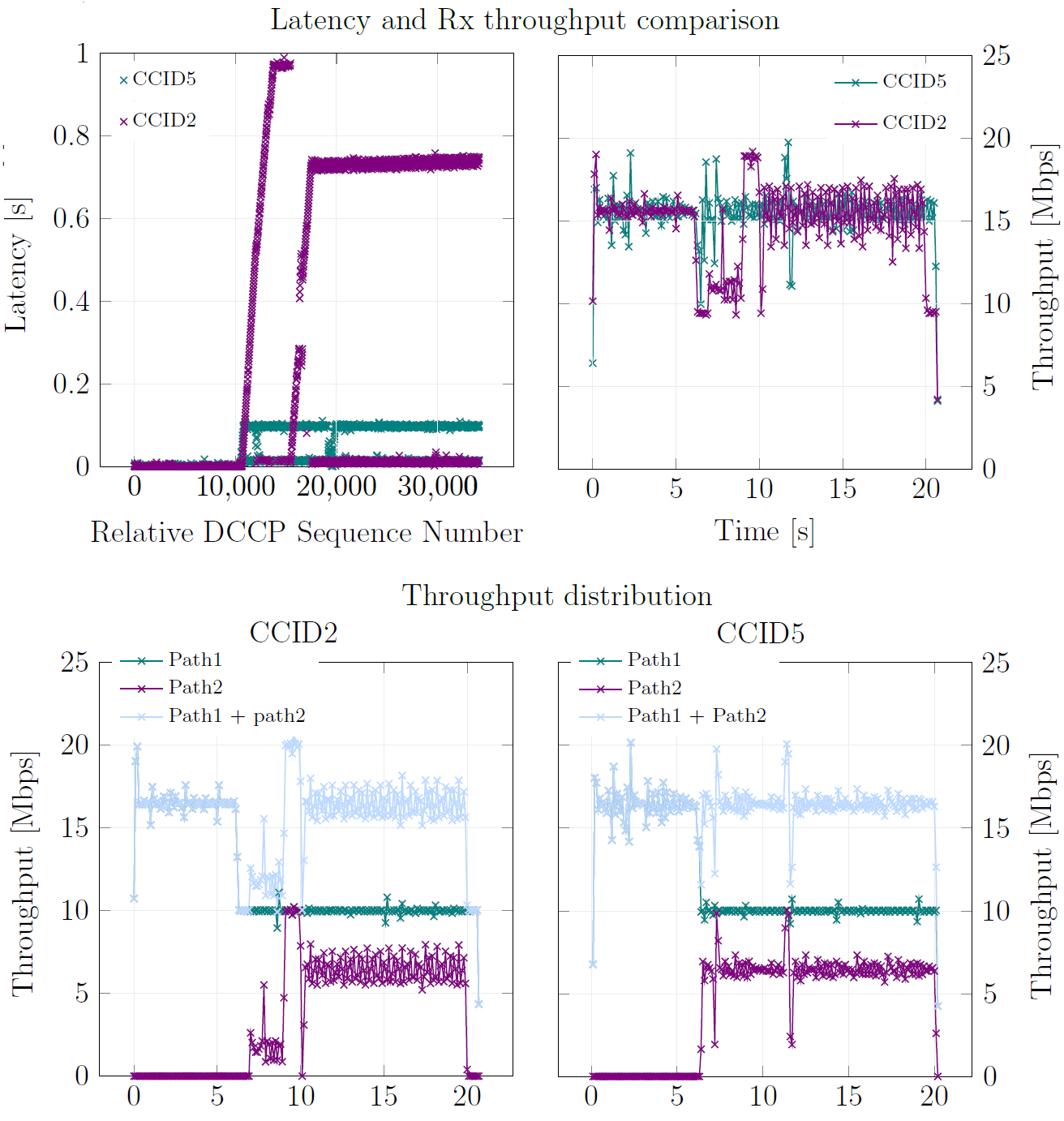}
	\caption{UDP over MP-DCCP - Latency, throughput and throughput distribution for CCID2 vs CCID5.}
	\label{fig:mp-udp-bwch}
	\Description{Throughput and Latency plots comparing CCID2 and CCID5 on UDP multi-path.}
\end{figure}

Figures ~\ref{fig:mp-udp-bwch} shows the results for the test described in ~\ref{mp-exp1} under the bandwidth change condition stated in Table ~\ref{tab:multi_udp_test2}. As expected, during the first 5 seconds both CC algorithms have approximately the same response and only one of the paths is utilized due to the priorization given by  the CPF scheduler. However, once the bandwidth change occurs, the differences in the performance become visible, mostly reflected in the end to end latency. When CCID2 is in place, the buffer created with the \texttt{tc} configuration prevents the CC algorithm to detect the bandwidth change and to report the congestion status to the scheduler. Thus, as the packets start to queue, the latency starts to increase as well. Once the buffer is completely full, a packet loss occurs, the \texttt{cwnd} is reduced and as a consequence the latency values drop. At this point, the scheduler gets the information of the congested path (near to t=8s), and starts allocating traffic to the second one. At the end, the primary path becomes fully utilized with a 10 Mbps throughput and the secondary path transports the remaining traffic to achieve the 15 Mbps imposed by the application. On the latency plot it is also possible to differentiate the latencies of both paths. On the other hand, when CCID5 is active, the bandwidth change is rapidly detected, causing the scheduler to react faster and  starting to transmit over the secondary path. The throughput is also distributed as it was with CCID2, 10Mbps over the primary path and 5Mbps over the secondary one. The latency difference between the primary and the secondary path can also be observed here, although the values for the congested link are lower in comparison with CCID2. The aggregated throughput is slightly higher than 15Mbs due to the overhead caused by the DCCP encapsulation process.
\subsection{Multi path scenario - TCP traffic}
\begin{figure}[h]
	\centering
	\includegraphics[width=\linewidth]{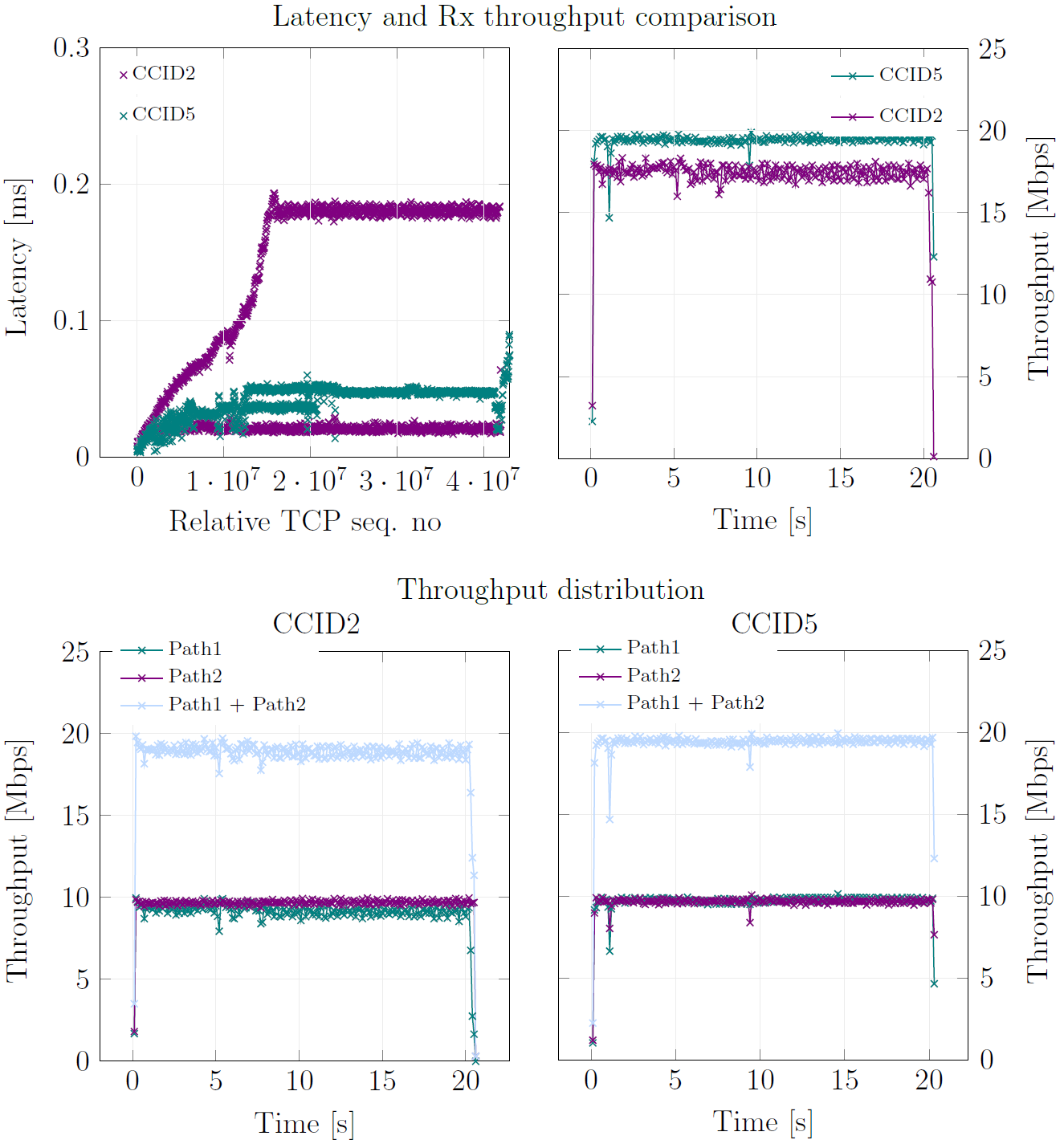}
	\caption{TCP over MP-DCCP - Latency, throughput and throughput distribution for CCID2 vs CCID5}
	\label{fig:mp-tcp-gen}
	\Description{Throughput and Latency plots comparing CCID2 and CCID5 on TCP multi-path.}
\end{figure}

The comparison of throughput and latencies measured for CCID2 and CCID5 is illustrated in
Figure~\ref{fig:mp-tcp-gen}. Both CC algorithms reach a throughput of 17.4 Mbps and 18 Mbps in average, which correspond to values close to the total BW available among both existing paths (20 Mbps).In terms of latency, CCID2 shows values concentrated around 0.02s and 0.18s, which shows that in this case, despite the attempt of the round robin mechanism to distribute the load equally among both paths, path2 stabilizes its cwnd at a larger value in comparison to path1, receiving more data and producing higher latencies. This difference in the saturation of both paths for CCID2 can be also observed in the throughput distribution graph from the same Figure~\ref{fig:mp-tcp-gen} , where path1 reaches a slightly lower throughput than path2. On the other hand, CCID5 shows latency values that have an increase from approx. 0.01s and 0.055s during the first half of the test, to finally get stabilized at a value of around 0.045s. This allows us to infer that the CCID5 algorithm managed to estimate properly the network conditions of both paths, leading to an equal distribution of the load and in consequence steady values of both the latencies and the effective throughput.

\subsection{Evaluation Summary}
The  results gathered under the single-path scenario show that the conceptual foundation of BBR, which was initially intended for TCP, is also extensible to DCCP, showing comparable responses in terms of throughput and latency. The sender works at its optimal operation point, where the amount of data in flight equals BDP, intermediate buffers are not filled up and the values of latency are lower than the ones achieved in CCID2.

In the multi-path scenario, when transporting UDP traffic, the results allow to conclude that the use of CCID5 has clear benefits with respect to CCID2 in the case of a bandwidth drop. Here, the rapid estimation of the new network conditions performed by CCID5  and the corresponding convergence of the sending behavior after the change detection, lead to a faster and better response of the CPF scheduling algorithm. At the end, the benefit is reflected in the achievement of a more steady throughput. In the case of TCP, the accurate estimation of the path bandwidth leads to an improvement in the load distribution performed by the Round Robin scheduler, thus, both paths get equally saturated and there is not a perceptible difference among their latencies. In addition, the latency values achieved in both cases (TCP and UDP transport) are lower in comparison with the ones obtained with CCID2, which also reduces the reordering efforts at the receiving end.

\section{Conclusion and Future work}
In this paper, we implemented BBR congestion control as a new CCID in the DCCP protocol, proving that the mechanisms proposed by BBR to characterize the network path and update the sending behavior accordingly, are suitable for DCCP under the hybrid access scenario. In addition, we integrated our approach into a multi-access framework within the 5G ATSSS context. This is the basis for future work, where we intend to evaluate additional variables such as: concurrent flows (fairness), path delay and packet loss. We also plan to bring BBRv2\cite{bbrv2} to the DCCP stack.  Version 2 of BBR seeks to improve fairness when co-existing with other CC algorithms like RENO or CUBIC, reduce loss rates, and use ECN signals for rate adaptation~\cite{bbrv2_slides}. Finally, specifying the BBR implementation for DCCP as part of its standardized CCID framework at IETF under a new CCID version 5 is another interesting path to follow.

\bibliographystyle{ACM-Reference-Format}
\bibliography{sample-sigconf}

\appendix
\end{document}